# Aggregation of ferromagnetic and paramagnetic atoms at edges of graphenes and graphite*

Zhou Hai-Qing (周海青)[a,c], Yang Huai-Chao(杨怀超)[a,c], Qiu Cai-Yu (邱彩玉) [a,c], Liu Zheng (刘政)[a,c], Yu Fang (余芳)[a,c], Hu Li-Jun (胡丽君)[a,c], Xia Xiao-Xiang (夏晓翔)[b], Yang Hai-Fang (杨海方)[b], Gu Chang-Zhi (顾长志)[b], and Sun Lian-Feng (孙连峰)[a] †

[a]National Centre for Nanoscience and Technology, Beijing 100190, China

[b]Institute of Physics, Chinese Academy of Sciences, Beijing 100190, China

[c]Graduate School of Chinese Academy of Sciences, Beijing 100049, China

†Corresponding author. E-mail: slf@nanoctr.cn

In this work, we report that when ferromagnetic metals (Fe, Co and Ni) are thermally evaporated onto n-layer graphenes and graphite, a metal nanowire and adjacent nanogaps can be found along the edges regardless of its zigzag or armchair structure. Similar features can also be observed for paramagnetic metals, such as Mn, Al and Pd. Meanwhile, metal nanowires and adjacent nanogaps can not be found for diamagnetic metals (Au and Ag). An external magnetic field during the evaporation of metals can make these unique features disappear for ferromagnetic and paramagnetic metal; and the morphologies of diamagnetic metal do not change after the application of an external magnetic field. We discuss the possible reasons for these novel and interesting results, which include possible one dimensional ferromagnets along the edge and edge-related binding energy.

Keywords: Graphene, edges of graphene, ferromagnetism, magnetic particle inspection (MPI)

## 1. Introduction

Experimental reports of ferromagnetism in graphite[1-5] and other carbon-based materials,[6-8] which are usually assumed to be typical diamagnetic materials, have attracted intense interests because only s/p electrons are present in contrast to traditional 3d/4f electrons. The extremely weak signals at room temperature lead to hot-debated origins of ferromagnetism in these carbon-based materials.[3,9,10] Meanwhile, monolayer graphene,[11] an individual atomic layer of graphite, with zigzag edges is predicted to give rise to magnetic ordering.[9,12] These predictions need convincing experimental verifications because Ising[13] and Heisenberg[14] models state that no long-range ferromagnetic order in one-dimensional system is possible at finite temperatures. In these spin lattice models,[13,14] short-range magnetic interactions are usually assumed and the interactions between one-dimensional structures with its surroundings are not taken into accounts. If interaction exists between monoatomic chains of Co on a Pt substrate, long range of ferromagnetic ordering in one-dimensional monatomic chains of cobalt atoms has been found.[15]

In this paper, we use a technique that is miniaturized one of the nondestructive testing magnetic particle inspection (MPI) widely used in industries. With this technique, the following results are obtained: 1) Nanowires and adjacent nanogaps can be found clearly at edges of graphenes and graphite for ferromagnetic and paramagnetic atoms (Ni, Fe, Co, Mn, Al, Pd). 2) The application of an external magnetic field has great effects on these features for ferromagnetic and paramagnetic metals; while the morphologies of diamagnetic metals (Au,Ag) do not change. We propose the possible reasons for these observations, which open up fields of future applications of carbon-based magnetism and spintronics.[16,17]

## 2. Experiment

N-layer graphenes were obtained from micromechanical cleavage of natural graphite (Alfa Aesar) and then transferred onto $SiO_2$/Si substrates.[18] After the layer number of n-layer graphenes was determined by optical microscope (Leica DM 4000) and Raman spectroscopy (Renishaw inVia Raman Spectroscope), a thin metallic film of Ni (Fe, Co, Mn, Al, Pd, Au, Ag) was evaporated onto the wafer in a vacuum thermal evaporator[18] without or with an external magnetic field produced by a permanent NdFeB magnet. When a permanent magnet was applied, the wafer and the magnet were amounted on an aluminum stage with the graphene at a position of known magnetic field. Finally, scanning electron microscope (SEM) was used to characterize the film morphologies at the edges of n-layer graphenes.

## 3.  Results and discussion

### 3.1. Special features of nickel film at edges of graphenes

The morphologies of nickel film on n-layer graphenes are closely related to the layer number of graphenes,[19] which can be used to identify the layer number.[18] After thermal deposition of nickel film onto n-layer graphenes (Fig. 1), the morphologies of nickel film on the central area of n-layer graphenes are distinctly dependent on the layer number, which is consistent with our previously reported results.[18]

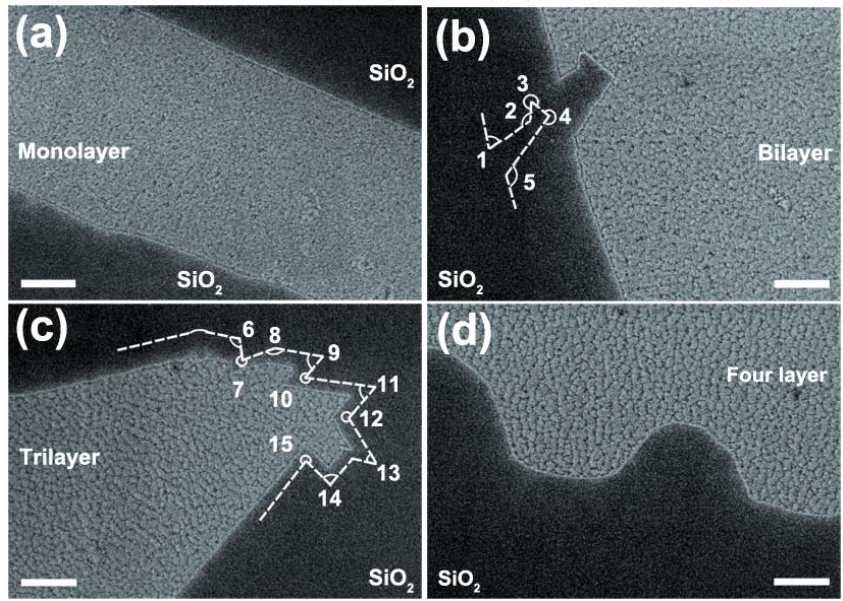

**Fig. 1.** SEM images of Ni nanowires, adjacent nanogaps at the edges of graphenes. Film thickness: 2.0 nm. Scale bar: 200 nm. (a) Monolayer graphene and substrate ($SiO_2$). (b) Bilayer graphene and substrate. (c) Trilayer graphene and substrate. (d) Four layer graphene and substrate.

Quite interesting, different features can be seen clearly in Fig. 1: there is a nickel nanowire along the edges of n-layer graphenes (n=1, 2, 3, 4), which is separated from the central nickel film by a clear nanogap. These features can also be found at the edges of n-layer graphenes no matter what kinds of the edges are: straight, curved or at angles (Fig. 1(a)-(d)), or at the edge of graphite. For example, the angles in Fig. 1(b) are $65.3^{\circ}$("1"), $127.3^{\circ}$("2"), $307.6^{\circ}$("3"), $262.0^{\circ}$("4"), $127.9^{\circ}$("5"), respectively; while the angles in Fig. 1(c) are $111.0^{\circ}$("6"), $285.8^{\circ}$("7"), $151.7^{\circ}$("8"), $60.0^{\circ}$("9"), $302.6^{\circ}$("10"), $56.9^{\circ}$("11"), $248.8^{\circ}$("12"), $46.1^{\circ}$("13"), $85.6^{\circ}$("14"), $276.6^{\circ}$("15"), respectively.

### 3.2. Paramagnetic and diamagnetic metal films on graphenes

The thickness-dependent morphologies of nickel on the central area of n-layer graphenes can be well explained from thermodynamic (e.g. energetics and stability) and kinetic (e.g. surface diffusion) factors as in the situation of gold.[18,19] Why the morphologies of nickel at the edges of graphenes are so unique and different? To have a better understanding of the mechanism, controlled experiments with other metal elements have been carried out, such as Fe, Co (ferromagnetic elements), Mn, Pd, Al (paramagnetic metals), Au and Ag (diamagnetic metals), respectively.

For Fe and Co, similar features can also be observed at edges of n-layer graphenes or graphite, which are not shown here. Meanwhile, metallic nanowires and adjacent nanogaps can also be found for Mn, Pd, Al atoms at edges of graphenes (Fig. 2).

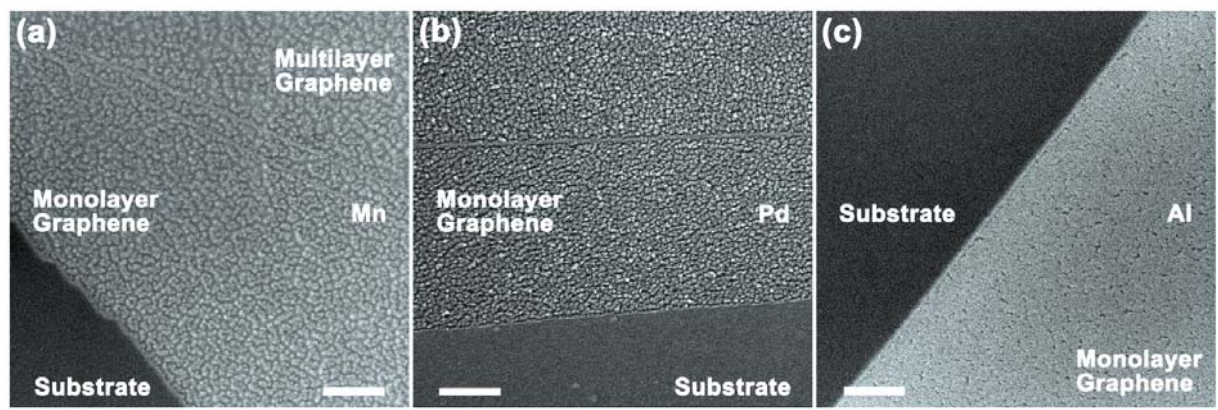

**Fig. 2.** SEM images of Mn, Al and Pd nanowires, adjacent nanogaps at the edges of n-layer graphenes or at terraced edges where graphenes stack together: (a) Mn; (b) Pd; (c) Al. Film thickness: 2.0 nm. Scale bar: 200 nm.

However, for diamagnetic elements (Au, Ag), the atoms tend to form isolated nanocrystals that are not connected to each other at the edges of graphenes when the thickness of metal (2.0 nm) is small (Fig. 3(a) and (b)).[20] When there are more atoms evaporated and a uniform smooth film (5.0 nm) forms, no nanowires and adjacent nanogaps are found at edges of n-layer graphenes or at the terraced edges, indicating that the aggregation is not similar to that found for nickel (Fig. 1).

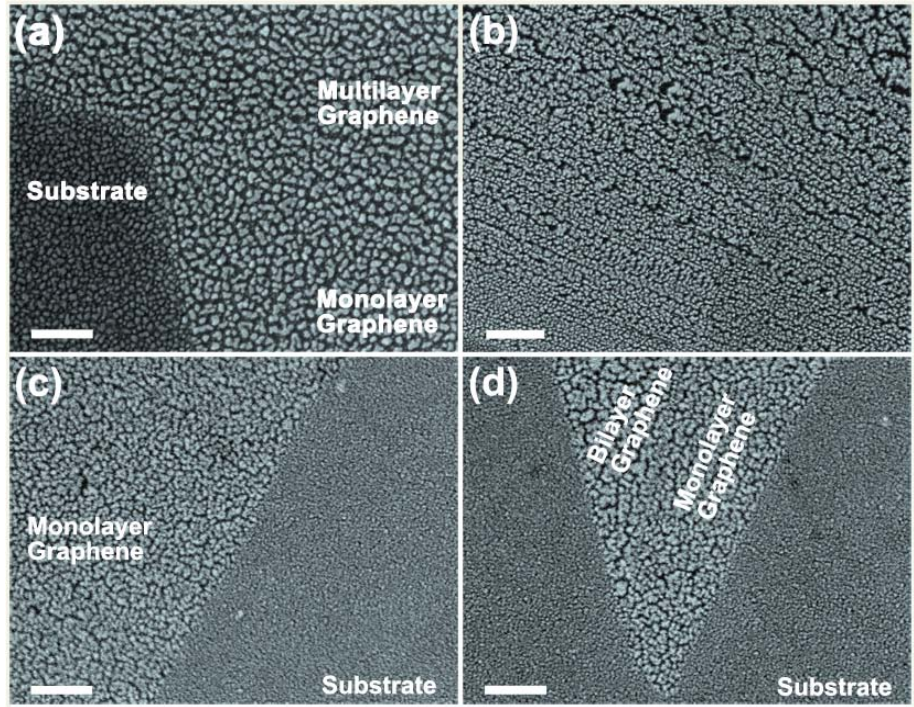

**Fig. 3.** SEM images of diamagnetic metals (Au and Ag) at the edges of n-layer graphenes or at terraced edges where graphenes stack together: (a) 2 nm Ag; (b) 2 nm Au; (c, d) 5 nm Au. Scale bar: 200 nm.

To figure out the possible mechanism of these features, an external magnetic field is applied during thermal deposition of 2 nm nickel or gold films onto monolayer graphene (Fig. 4). As is shown in Fig. 4 (b), when an external field (~0.34 Tesla) is applied parallel to the surface of monolayer graphene, the morphologies of 2 nm gold film almost have no changes. However, if an external magnetic field (~0.35 Tesla) parallel to the graphene surface is applied (Fig. 4(d)), the nickel nanowire/nanogap disappears at the edge of monolayer graphene, and the nickel film at the central region changes significantly. These results indicate the great effect of external magnetic field on the morphologies of nickel film on graphenes.

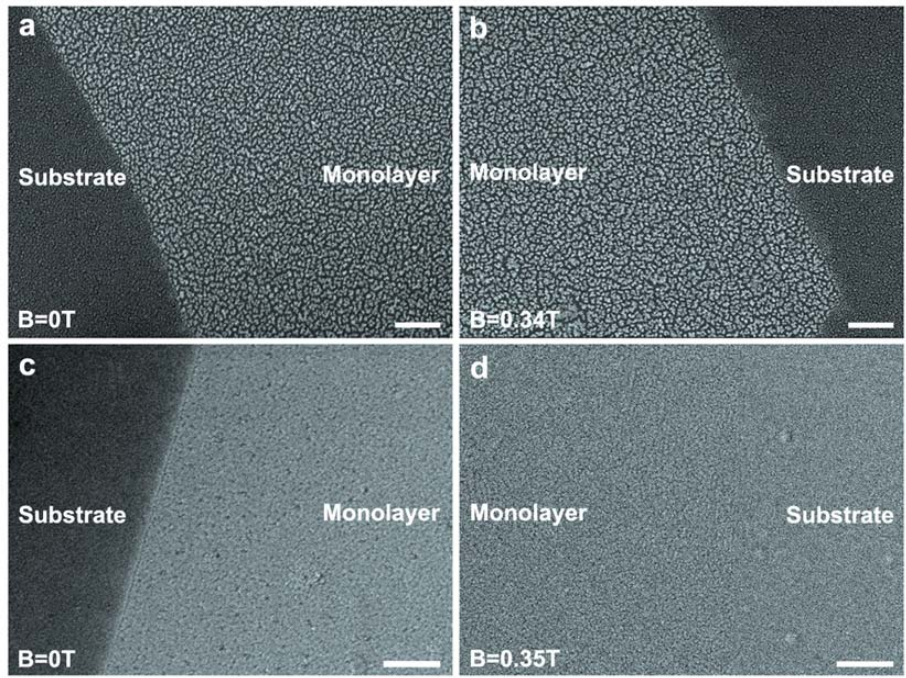

**Fig. 4.** SEM images of 2.0 nm Au or Ni film on the surface of monolayer graphenes with and without external magnetic field at 25°C. Scale bar: 200 nm. (a) Gold film without external magnetic field. (b) Gold film with an external field. Direction of magnetic field: left to right. (c) Nickel film without external magnetic field. (d) Nickel film with an external field. Direction of magnetic field: left to right.

### 3.3. The possible mechanisms

What are the mechanisms of these unique features found at the edges of graphenes?

One possible reason is the short-ranged, strong chemical forces at the edges of graphenes.[18,19,21] The metal atoms of (Ni, Fe, Co, Mn, Al and Pd) may show higher binding energies at the edges than that for the interior binding sites. Although this chemical force may play roles in the process of atom diffusion at the edges, it can be excluded as the reason for the formation of nanowire and nearby nanogap due to the following reasons: firstly, the degree of a chemical binding site will show saturation. This implies that the chemical force only affects a very small and limited number of metal atoms at the edges. For example, in 2-nm-thick films (around 10 monolayers), about 10% of atoms are at the interface with graphenes. That is to say that 90% of the atoms are not affected by the possible short, strong chemical bonds. Secondly, if chemical binding is the reason for the nanowire and adjacent nanogap at the edges of graphenes, an external magnetic field (~0.35 Tesla) will not change these features. This is because the changes of magnetic potential energy under the external magnetic field are too small (~$2\times10^{-5}$ *eV* for a Bohr magneton) to affect the binding state. Morphologies of gold film on graphenes are found to be unchanged after the application of an external magnetic field (Fig. 4(a) and (b)). Meanwhile, the morphologies of Ni film have changed significantly under the external field (Fig. 4(c) and (d)).

Another mechanism is similar to that in magnetic particle inspection (MPI) technique, in which the aggregation of magnetic particles indicates the existence of a nonuniform magnetic field. Here the difference is that metal atoms with magnetic moment are used instead of magnetic particles since local magnetic field will attract these atoms no matter they are ferromagnetic or paramagnetic. Therefore, the observations (nanowire and adjacent nanogap) can be attributed to the attraction and aggregation of atoms with magnetic moments, which indicates the existence of intrinsic magnetic moments at the edges of graphenes. According to this model, formation of nanowire and adjacent nanogap can be explained as the followings: when atoms with magnetic moments are thermally deposited onto the boundary area between n-layer graphenes and substrate ($SiO_2$), these adatoms will be attracted by the nonuniform magnetic field produced by the magnetic moments at the edges of graphenes. Thus, a metal nanowire and an adjacent nanogap can be observed at and on the edges of graphenes under SEM (Fig. 1 and 2). For terraced edges between graphenes, the magnets at the edge will attract atoms from both sides, which results in a nanowire and two adjacent nanogaps at these edges (Fig. 2(a) and (b)). The formation of nanowires and adjacent nanogaps at straight or curved edges indicate that the orientation of the local magnetic moments at the edges should be in the plane of n-layer graphenes.

Another interesting, unexpected observation is the existence of ferromagnetic ordering at the edges of graphenes no matter what kinds of the edges are: straight, curved or at angles (Fig. 1). Theoretical calculations predict that ferromagnetism may exist only in graphenes with zigzag edges.[9,12] It is not possible for all the edges are zigzag because there are angles of odd multiples of $30^o$ (Fig. 1), which indicates the two edges of these angles have different chiralities, i.e. one armchair and the other zigzag.[22,23] This contradiction may be resolved in this way: theoretical calculations usually assume ideal graphene edge structures: armchair or zigzag. In experiments, however, the edges of graphene may exist atomic roughness. For example, there are experimental works published recently,[24,25] which indicate the possibility of irregular edge structure of graphenes. The correlation of edge structure with ferromagnetism needs more theoretical and experimental works for future studies.

We hope to point out that the extension of the MPI to nanoscale is valid although the magnetic potential energy changes in a magnetic field experienced by an adatom (~ 1 Bohr magneton) on graphenes is about 3 orders of magnitude below KT at room temperature. Firstly, when adatoms diffuse on graphenes which is thermally activated, the motion is two-dimensional, random and nondirectional. A nonuniform magnetic field can make the adatoms have a directional movement even if the magnetic potential energy changes are small. The results of this directional movement are the accumulation and aggregation of adatoms at the edges of graphenes. Secondly, the force on a magnetic moment by a nonuniform magnetic field is proportional to the gradient of this field. For the applied external magnetic field by the NdFeB magnet, the gradient is about 35 T/m. This magnetic field has significant effect on the process of diffusion, aggregation of adatoms and the final morphologies of metal film.

## 4. Summary

In this work, ferromagnetic (Fe, Co and Ni), paramagnetic (Mn, Pd and Al) and diamagnetic (Au and Ag) metals are thermally deposited onto n-layer graphenes and graphite. For ferromagnetic and paramagnetic meatals, the aggregation and adjacent nanogaps are found along the edges of n-layer graphenes and graphite. With the application of an external magnetic field, the nanowire/nanogap features disappear at the edge of monolayer graphenes for atoms with magnetic moments together with the change of film morphologies at the central regions. These results suggest the ferromagnetic ordering at the edges of graphenes and graphite, which deserve more experimental and theoretical studies.

**Acknowledgements**

This work was supported by National Science Foundation of China (Grant Nos. 10774032, 90921001, 50952009).

# Aggregation of ferromagnetic and paramagnetic atoms at edges of graphenes and graphite*

Zhou Hai-Qing (周海青), Yang Huai-Chao (杨怀超), Qiu Cai-Yu (邱彩玉), Liu Zheng (刘政), Yu

Fang (余芳), Hu Li-Jun (胡丽君), Xia Xiao-Xiang (夏晓翔), Yang Hai-Fang (杨海方),

Gu Chang-Zhi (顾长志), and Sun Lian-Feng (孙连峰)†

　　本文中，通过在石墨烯或石墨上热蒸镀铁磁性金属 (铁,钴,镍),我们发现在石墨烯或石墨的边缘形成一条金属纳米线和相邻的纳米间隙。这种现象与边缘齿式或摇椅式结构无关。同时，这种现象存在于锰，钯，铝等顺磁性金属，但在抗磁性金属(金和银) 却没有发现。如果蒸镀金属时，沿石墨烯的表面方向上外加磁场，这种纳米线/纳米间隙结构将消失。然而，对于抗磁性金属Au或Ag，即使有外加磁场，热蒸镀金属的形貌以及边缘分布没有变化。我们讨论了可能的机理，提出实验显示石墨烯（石墨）边缘处存在本征磁矩或磁有序。